# Direct (3+1)D laser writing of graded-index optical elements

Xavier Porte,[1,†,*] Niyazi Ulas Dinc,[2,3,†,*] Johnny Moughames,[1] Giulia Panusa,[2]

Caroline Juliano,[2] Muamer Kadic,[1] Christophe Moser,[3] Daniel Brunner,[1] Demetri Psaltis[2]

[1]*Institut FEMTO-ST, Université Bourgogne Franche-Comté CNRS UMR 6174, Besançon, France*
[2]*Optics Laboratory, École Polytechnique Fédérale de Lausanne, Lausanne, Switzerland*
[3]*Laboratory of Applied Photonics Devices, École Polytechnique Fédérale de Lausanne, Lausanne, Switzerland*
†*Equally contributing authors*

*Corresponding authors: javier.porte@femto-st.fr ; niyazi.dinc@epfl.ch

**We propose single-step additive fabrication of graded-index optical elements by introducing the light exposure as the additional dimension to three-dimensional (3D) laser writing, hence (3+1)D writing. We use a commercial printer and photoresist to realize the proposed single step fabrication method that can be swiftly adopted for research and engineering. After presenting the characterization of the graded-index profiles via basic structures, we demonstrate two different optical devices: volume holograms that are superimposed using angular and peristrophic multiplexing, and optical waveguides with well-defined refractive index profiles. In the latter, we precisely control the propagating modes via tuning the (3+1)D-printed waveguide parameters and report step-index and graded-index core-cladding transitions.**

## 1. INTRODUCTION

Additive manufacturing by two-photon polymerization (TPP) is rapidly becoming an important hybridization tool among various photonic integrated platforms due to its high versatility. Neither photo-masks nor etchings are required for additive manufacturing, and intricate dielectric photonic structures can accommodate a wide range of host platforms, for example quantum systems [1], silicon photonics [2] or the tip of fibers [3]. Crucially, this fabrication technology is largely agnostic to particularities of such host platforms. Sub-wavelength feature sizes, i.e. beyond diffraction limit resolution, makes TPP highly attractive for meta-optics [4], high density photonic integration [5,6] as well as for functionalization via free-form optical components [7,8] and multilayer diffractive elements [9]. These applications often require intricate designs, and the precise control of analogue 3D refractive index distributions in 3D additive manufacturing is highly desirable. Besides approaches leveraging binary index modifications of a single host material, there are multi material approaches to obtain additional degrees of freedom besides geometry. Sequentially introducing different photoresists through microfluidic channels has for example been reported [10,11]. Two-component photopolymers present high dynamic range alternatives [12-14] and multiple patterning on these materials for even more dynamic range has been demonstrated [15]. Recently, graded-index (GRIN) lenses relying on varying polymerization inside a porous host material have been demonstrated [16]. In addition, fabrication of a biological cell phantom via TPP that has multiple refractive index values by varying polymerization has been reported [17].

In this work, we rely only on a single resin to create GRIN optical volume elements in a single step with a commercial 3D printing machine by having the exposure as the fourth variable in the writing process; hence establishing (3+1)D printing to access and modify each voxel's refractive index independently. For that purpose, we rely on the exposure-dependent refractive index of broadly utilized commercial photoresists [18-21], dynamically modifying the exposure during the printing process to construct optical waveguides and volume holograms with single-step, single-material, and a commercially available process.

The waveguide is a principal component of integrated photonics, as evidenced through applications like photonic wire bonding [22,23], on-fiber direct laser writing [3, 24] and in the context of scalable photonic interconnects [5,6]. Core-cladding 3D-laser printed structures guide light with a refractive index of the core moderately larger than that of the cladding. Such waveguides are currently the subject of intense study, and basic photonic tools like single-mode low-loss waveguides [25] or multimode light splitters [26] are being demonstrated. However, those either require direct inscription into a bulk material [16], intricate 3D photonic crystal fiber structures [27], or a multi-step process relying on different materials to provide the required refractive index difference between core and cladding [2,28]. Our approach based on (3+1)D printing requires a single fabrication step only and enables precise control of the refractive index's profile.

Another fundamental and highly relevant 3D structure is the volume hologram [29-31]. A stack of 2D images can be multiplexed in a volume hologram and each of these images can be retrieved individually by employing different angles, wavelengths or phase distributions. In a volume hologram, each 2D data corresponds to an individual hologram to be multiplexed with several others, and the high resolution of TPP means the technique is excellently suited for their fabrication. Crucially, since TPP provides independent access to each voxel in the fabrication volume, a hologram can be designed digitally without considering a recording schedule and avoiding the crosstalk between the individual recording sequences.

Figure 1 illustrates our GRIN printing concept for the two applications discussed here: volume holograms (Fig. 1(a)) and photonic waveguides (Fig. 1(b)). Leveraging such (3+1)D, or gray-tone lithography, our technique elegantly capitalizes on available and established 3D TPP materials and equipment and can be widely adopted without delay.

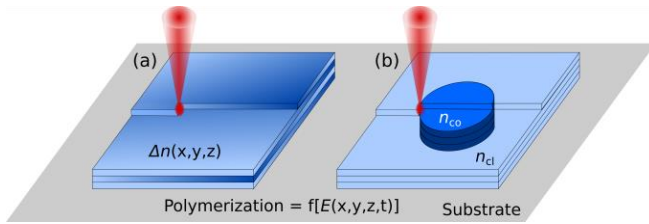

**Fig. 1.** Concept of (3+1)D printing: the refractive index ($\Delta n$) is controlled by dynamically changing the laser power during the printing process ($E$(x,y,z,t)). Illustration of this concept in a (a) volume grating used for holography and a (b) photonic waveguide with a step-transition between refractive indices of the central core ($n_{co}$) and the surrounding cladding ($n_{cl}$), printed with respective laser beam field amplitudes $E_{co} > E_{cl}$.

## 2. FABRICATION METHODOLOGY

The Clausius–Mossotti relation states that the refractive index of a material depends on the number of molecules per unit volume [32, 33]. Many photo-resins yield greater density in the polymer phase than in the resin phase, and the degree of polymerization depends on the polymerization kinetics, which is driven by light intensity and exposure time. Thus, the refractive index can be modulated by changing the light exposure [18-21], which is the here leveraged physical mechanism to demonstrate GRIN structures printed with the (3+1)D principle. We used commercial direct-laser writing systems from Nanoscribe GmbH (Photonic Professional GT+) and the negative tone photoresist "IP-Dip".

The (3+1)D printing process can be dissected in the following steps: the photoresist is dropped on a fused silica glass substrate (25 x 25 x 0.7 mm³) and is photo-polymerized via TPP with a 780 nm femtosecond pulsed laser, focused by a 63X (1.4 NA) microscope objective. The printing process was realized in consecutive horizontal layers as schematically illustrated in Fig. 1, and the laser power was dynamically modified for each voxel to produce the desired 3D refractive index distribution. The vertical (slicing) and the horizontal (hatching) sampling resolution is 0.3 μm and 0.1 μm, respectively. For volume holograms, the scan speed of the writing beam was chosen to be 7 mm/s and the laser power range is set to be 28-42% of the maximum average power that is 20 mW, while for photonic waveguides, scanning speed was chosen to be 10 mm/s and the laser power range was set to be 35-58% of the maximum average power. Therefore, the utilized laser fluences (J/cm²) per unit volume are comparable for both applications, once normalized with respect to the scan speed. The particular values are the result of a careful optimization, balancing the refractive index uniformity and mechanical stability. Very low degrees of polymerization result in unstable structures that do not hold the intended shape, while very high degrees of polymerization are unavailable due to burning and micro-explosions occurring during writing.

We used an off-axis digital holographic interferometry (DHI) setup for the principle identification of conditions resulting in good structure-refractive index fidelity (see Supplementary Discussion 1 for further details). After printing and during the development process the non-polymerized leftover resin is washed away for the fabrication of the waveguides. The development process results in a small degree of structural shrinkage, where in particular volumes with a relatively lower degree of polymerization are affected most. As waveguides are embedded in a comparably large surrounding volume, structural stability is inherently given and shrinkage is no issue. However, the development step is skipped for volume gratings and holograms so that shrinkage that is occurring during development is avoided, and furthermore that the liquid resin serves as an index matching liquid. For high exposure values, the polymerization yield is complete and the material is mechanically stable [34].

## 3. CHARACTERIZATION OF THE GRIN PROFILES

To calibrate the laser power vs refractive index relation, a rectangular prism is printed with the power increasing linearly along one transverse dimension. The phase accumulation through this structure is measured by DHI. Knowing the thickness, the measured phase is converted to a refractive index difference with respect to the background, which is the not polymerized resin. The obtained index difference is given in Fig. 2(a). An exponential curve is fitted on the experimental data, which we use for mapping a desired refractive index to the corresponding writing power. Taking this nonlinear dependency into account is particularly crucial for writing gratings and holograms, since an accurate refractive index distribution is essential for maximum diffraction efficiency. Hence, the laser power is adjusted according to the target refractive index value at each voxel position.

In Fig. 2 (b), we provide the mean and standard deviation of this measurement obtained from five different samples fabricated with the same writing scheme and parameters. This averaged measurement results in relatively large error bars, especially towards the low index difference. The first reason, which applies equal to all the data points, is the noise level in the digital interferometry setup for the phase measurements. The second reason are small sample to sample variations in the laser writing caused by allocation of the sample interface. Optical resolution limits the accuracy of allocating this interface to a few tens of nanometers, which in turn leads to different phase accumulation along the thickness of the sample and hence a systematic offset variation for each sample. On top of that, slightly different focusing in DHI for different samples contributes to sample-to-sample differences. These effects become relatively stronger for the regions where the refractive index change is small (see Supplementary Discussion 2 for further details). Since the relative index change within a sample is more meaningful, we stick with the calibration curve and following power adjustment obtained from a single sample. Nonetheless, we provide the result by averaging over different samples as a general reference, as it reports relevant information about the fabrication process. The significant effect of the nonlinear power vs refractive index relationship is shown in Fig. 2(c), where phase extractions are given from sinusoidal gratings that were printed assuming a linear power dependence or the exponential fit to experimental data. The arrows in the zoom-in panels of Fig. 2(c) highlight distortions on the sine wave due to the saturation of refractive index (or polymerization) at high power caused by the linear approximation.

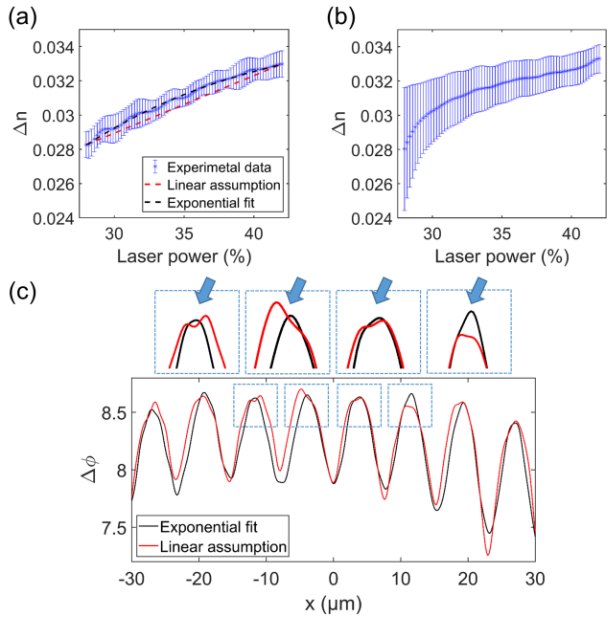

**Fig. 2.** (a) Refractive index difference of the polymerized part with respect to the monomer resin vs laser power from a sample. (b) Refractive index difference of the polymerized part with respect to the monomer resin vs laser power averaged from five samples. (c) Phase extractions from the 20-μm thick sinusoidal gratings printed with linear power dependence assumption and the exponential fit where the arrows highlight the effect of saturation.

To cross-validate the obtained results from DHI experiments, we printed several volume gratings. A volume grating is a sinusoidal refractive index distribution in any orientation. Therefore, any arbitrary distribution can be represented by a superposition of volume gratings, which makes it the fundamental 3D building block of GRIN volume elements. We printed unslanted volume gratings with a 6 μm period, which yields an approximately 3.2° Bragg angle at 1.03 μm wavelength in 1.52 refractive index. The transverse dimension is (120 μm)² and the thickness is varied from 30 μm to 120 μm in 30 μm steps. High dynamic range and sample thickness lead to over-modulated gratings where the diffraction efficiency at the Bragg condition decreases and the energy goes to the lateral lobes in the angular spectrum [35]. Gratings thicker than 120 μm are not investigated to keep the samples in the under-modulated regime. The calibration curve shown in Fig. 2(a) is utilized to fabricate the sinusoidal GRIN profile as depicted in Fig. 3(a), where all the volume is polymerized with different degrees of polymerization by (3+1)D printing and targeting a dynamic range of $\Delta n = 5 \cdot 10^{-3}$. Fig. 3(a) shows the 90-μm thick grating where the z-axis is the optical axis. All the volume gratings are printed on the same substrate with 250 μm center-to-center separation. The first, 30 μm thick grating takes 40 minutes to fabricate, which is the same for uniform exposure of an equivalent structure in terms of size, sampling, scanning speed; varying exposure does therefore not impact fabrication time, and fabrication time only scales linearly with the printed volume.

A collimated beam with a dimension that is comparable to the transverse area of a single volume grating is used as the input beam. This input beam illuminated each volume grating one by one, and the transmitted beam is recorded while the input angle is varied to measure Bragg selectivity. The efficiency vs thickness curve is given in Fig. 3(b). By using the coupled wave theory pioneered by Kogelnik [36, 37], $\sin^2$ curve fitting is performed. From the argument of the $\sin^2$ fit, $\Delta n$ is found to be $5.1 \cdot 10^{-3}$, which is in excellent agreement with the targeted dynamic range and hence an independent confirmation of the rectangular prism-based calibration via DHI. The angles of the input plane waves are swept around the Bragg angle for all samples of different height, and the efficiency relation by [36, 37] is employed for curve fitting as exemplified in Fig. 3(c) for the 90 μm thick grating. The first null is labeled as $(\Delta\theta)_B$ referring to Bragg selectivity. Figure 3(d) shows the Bragg selectivity for all thickness values along with the theoretical curve that is numerically calculated for $\Delta n = 5.1 \cdot 10^{-3}$, where we observe a good match with the experimental results (see Supplementary Discussion 3 for further details).

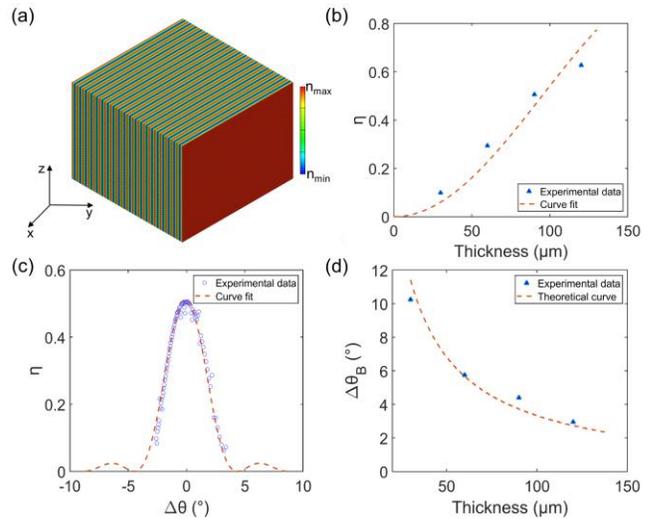

**Fig. 3.** (a) Visualization of the printed volume gratings, (b) the efficiency vs the structure thickness plot, (c) the Bragg selectivity measurement around the Bragg angle for the grating of 90-μm thickness, (d) the Bragg selectivity vs the structure thickness plot of the experimental data along with the numerically calculated theoretical curve.

Finally, the optical response of (3+1)D printed samples does not significantly change over time, indicating that there is negligible aging effect, which is in accordance with the results reported in [19], while the final density of the material can be subject to small changes due to diffusion of monomers and post-polymerization. In Supplementary Discussion 4, we provide DHI measurements of the same phase grating on various days after fabrication, where we do not observe any difference larger than experimental error range. The same finding was confirmed independently based on our waveguides, which did not experience transmission or confinement modifications over a time-span exceeding one year.

## 4. GRIN OPTICAL ELEMENTS

### a. Volume holograms

Volume holograms have been of great interest for parallel optical interconnects [38], data storage [39] and optical mode multiplexing and demultiplexing [9,40]. TPP has also been identified as a candidate technique for optical data storage [41]. When a hologram is optically recorded in a photosensitive medium by interfering a reference ($E_{\text{ref}}$) and object ($E_{\text{obj}}$) beam, the recorded hologram is given by:

$$|E_{\text{ref}} + E_{\text{obj}}|^2 = E_{\text{ref}}E_{\text{ref}}^* + E_{\text{obj}}E_{\text{obj}}^* + E_{\text{ref}}E_{\text{obj}}^* + E_{\text{ref}}^*E_{\text{obj}} \quad (1)$$

The last term in Eq. (1) provides the reconstruction of the object beam upon illumination with the reference. The first two terms constitute the DC component and the third term is the conjugate of the reconstruction term. For a linear and reciprocal polymer media, the refractive index modulation follows Eq. (1). As a result, we can state that the object beam is a signal that is modulated by a carrier generated by the reference beam. To multiplex M holograms, we can use M different object beams and slightly different carrier frequencies, which are sufficient to be individually retrieved in a volume hologram by taking advantage of the angular Bragg selectivity. When M holograms are stored, the efficiency of an individual hologram is inversely proportional with M squared, which is a well-established rule for both photorefractive and photopolymer media [42-45]. For a photopolymer, while multiplexing many holograms, the DC terms build up within the dynamic range of the material, leaving only a fraction to the signal terms. This phenomenon decreases the holograms' diffraction efficiency since the diffraction efficiency of each hologram is related to its share from the available dynamic range. When the refractive index change in a recording media is not bipolar, the DC build-up cannot be removed.

In contrast to classical optical recording, additive manufacturing enables updating the index value of a single voxel in the media without affecting the others. Hence, we can design an index distribution digitally and fabricate it thanks to voxel by voxel (3+1)D printing. On a computer, we can superimpose holograms by eliminating the DC terms as if negative intensity values and bipolar responses were available. The obtained refractive index distribution can then be scaled to match the dynamic range of the material, allowing fabrication via TPP leveraging only monotonic index change using physical (non-negative) intensity values, while importantly preventing the DC build-up.

To demonstrate this approach as a proof of concept, the initials of the Optics Laboratory, O and L, are chosen to be peristrophically multiplexed in a (50 μm)$^3$ volume as presented in Fig. 4(a). The images of the letters are propagated digitally by the Beam Propagation Method (BPM) and the diffracted fields are phase-conjugated and superimposed with two plane waves tilted at 6.5° in x and y axis for each letter, eliminating the DC terms. In Fig. 4(a), the refractive index distributions of the first (at z=0), middle (at z=25 μm) and last (at z=50 μm) layers are demonstrated, where the target refractive index dynamic range is $5 \cdot 10^{-3}$. A bright field microscope image of the sample is given in Fig. 4(b). The obtained experimental reconstructions are given in Fig. 4(c-d) where the volume hologram is illuminated with the assigned plane wave.

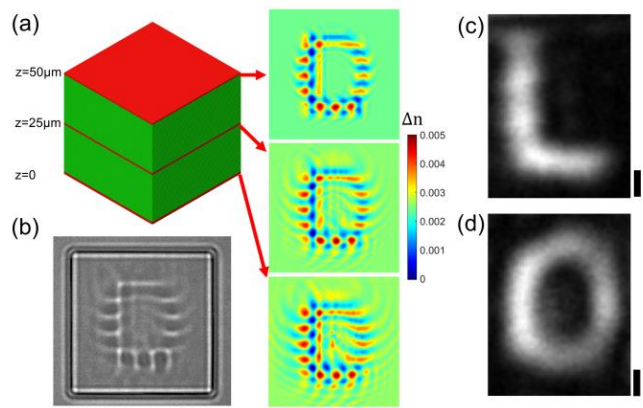

**Fig. 4.** (a) 3D rendering of volume hologram with the example index distributions in transverse planes at z=0, z=25 and z=50 μm, (b) bright field microscope image of the fabricated hologram whose transverse dimensions is 50x50 μm$^2$, (c-d) experimental reconstruction of letters 'L' and 'O'. The scale bars are 5 μm. All index distributions are up sampled for better visualization and share the same colorbar.

To demonstrate angular multiplexing as well, four arbitrarily chosen linearly polarized (LP) modes of a multimode fiber (LP$_{21}$, LP$_{12}$, LP$_{11}$, and LP$_{31}$) are multiplexed in a volume of 50x50x60 μm$^3$ following similar steps. Four carriers were chosen corresponding to 7° and 13° in both x and y-axes. The diffraction of the chosen LP modes can be neglected for 60 μm propagation distance since their spatial distributions have low frequency components. Hence, the z-variation is kept uniform for this hologram. The target XY refractive index variation is given in Fig. 5(a), the bright field microscope image of the sample is given in Fig. 5(b), and the obtained experimental intensity reconstructions are given in Fig. 5(c-f). The fabrication times for both holograms are roughly 150 minutes where it is 13 minutes for uniform exposure of an equivalent structure in terms of size, sampling, scanning speed. The reason for this discrepancy is software-related. For the volume gratings, we do not observe this discrepancy since the varying laser power is expressed with a simple formula, which is not possible for the holograms. An array should store the power values for each voxel, however, the DeScribe printing software does not allow for an array structure. Thus, power values are stored via thousands of if-else statements, which slows down the execution by control software considerably.

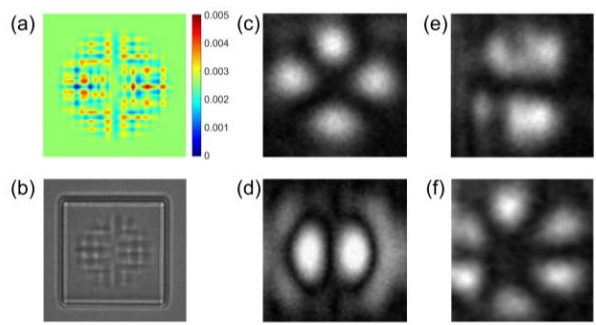

**Fig. 5.** (a) The target index distribution of the angular-peristrophic hologram, (b) bright field microscope image of the fabricated hologram whose transverse dimensions is 50x50 μm$^2$, (c-f) reconstructed images corresponding to LP$_{21}$, LP$_{12}$, LP$_{11}$, and LP$_{31}$ respectively. The index distribution is up sampled for better visualization.

## b. Photonic waveguides

We (3+1)D-printed photonic waveguides with both, step-index (STIN) and graded-index (GRIN) refractive index profiles. As schematically illustrated in Fig. 1(b), the waveguide's core printed with high laser power is surrounded by a cladding printed with a lower laser power. We used writing powers of 58% (11.6 mW) and 35% (7 mW) for the highest and lowest refractive index, respectively. STIN waveguides result from a constant laser writing power all across their core, while for the core of GRIN waveguides writing power changes from high to low along a parabolic profile. We printed 25 STIN waveguides with diameters $D \in [1 \dots 20]$ μm, and 13 GRIN waveguides with diameters $D \in [5 \dots 10]$ μm, all embedded in cuboids of 300 μm height. The scanning electron microscopy micrograph of Fig. 6(a) depicts an exemplary cuboid accommodating 20 printed STIN waveguides. The fabrication time for such a large structure is ~180 minutes. Noteworthy, this time can be drastically reduced by ~2/3 if the volume between waveguides that do not require (3+1)D printing is photo-polymerized by blanket UV exposure. By printing waveguides with different diameters $D$, we effectively scan their normalized frequency $V = \frac{\pi}{\lambda_0} D \cdot NA$, where $NA = \sqrt{n_1^2 - n_2^2}$ is the numerical aperture, $\lambda_0 \simeq 660$ nm is the wavelength of the illumination laser and $n_1$ ($n_2$) are the refractive indices of core (cladding).

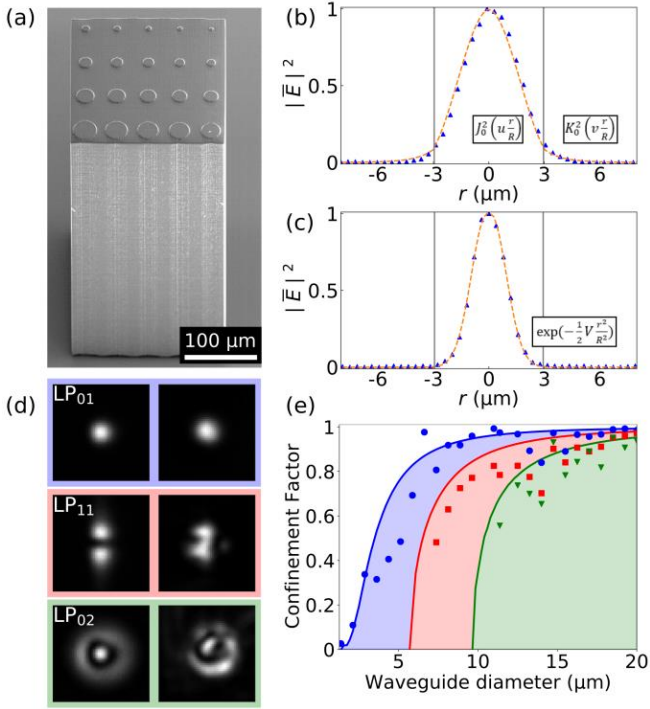

**Fig. 6.** (a) SEM image of a cuboid with 20 printed STIN waveguides. Panels (b) and (c) depict the output intensities (triangles) and fundamental mode fits (dashed lines) of a 6 μm diameter STIN and a GRIN waveguides, respectively. (d) Exemplary intensity profiles of injected LP modes (left) and their outputs after propagation through STIN waveguides of different diameters (right). We depict the first waveguide with confinement factor above 0.8 for each respective mode. (e) Theoretical confinement factor as function of the STIN waveguides diameter $\Gamma(D)$. Symbols are experimental and lines are theoretical curves for each mode $LP_{01}$ (blue), $LP_{11}$ (red), and $LP_{02}$ (green).

We extract the waveguide's relevant parameters by fitting the experimental output intensities for diameters below the cut-off condition of the second propagating mode. Figure 6(b) depicts the fit of $LP_{01}$ to the normalized output of a STIN waveguide with radius R = 3 μm, where the intensity profile of the $LP_{01}$ mode is given by $J_0^2\left(u\frac{r}{R}\right)$ for $|r| \leq R$ and $K_0^2\left(v\frac{r}{R}\right)$ for $|r| \geq R$. The output of a GRIN waveguide with same radius (3 μm) is shown in Fig. 6(c). The normalized experimental intensity is fitted with $\exp(-\frac{1}{2}V\frac{r^2}{R^2})$, which corresponds to the intensity distribution of the fundamental mode of a GRIN fiber with an infinite parabolic refractive index profile [46]. From the fit parameters we extract the corresponding NAs (see Supplementary Discussion 5 for further details). Using $n_1 = 1.547$ as the refractive index of IP-Dip at saturation [11], we obtain an averaged numerical aperture of NA = $0.08 \pm 0.01$ (i.e. $n_1 = n_2 + 2.4 \cdot 10^{-3}$) for STIN and of NA = $0.18 \pm 0.02$ for GRIN. Comparing the intensity profiles and the averaged NAs of single-mode GRIN and STIN waveguides we evidence that the core-confinement of the former is significantly higher, which offers a crucial advantage for photonic integration schemes [5,6].

In order to investigate the modal propagation properties, we used a spatial-light modulator (Santec LCOS SLM-200) to generate a set of LP modes, which we injected into the waveguides under close to NA-matched conditions (see Supplementary Discussion 6 for detailed description of the experimental setup). Figure 6(d) shows injected LP modes (left panels) and exemplary waveguide outputs (corresponding right panels). For each printed waveguide, we calculated the confinement factor $\Gamma(D)$ defined as the fractional optical power confined to the core. The experimentally measured confinement factors $\Gamma(D)$ for STIN waveguides are plotted in Fig. 6(e), together with the theoretically calculated curves for each mode. The experimental $\Gamma(D)$ follows the theoretical prediction with a small systematic vertical shift that we attribute to a systematic offset in the waveguides' diameter due to the non-negligible voxel dimensions.

We furthermore investigate the propagation-length dependent losses and the maximum waveguide packing density. Propagation losses primarily depend on scattering, host-material absorption and insufficient mode confinement. We determined the global losses of the $LP_{01}$ mode after propagating through the waveguides with lengths ranging from 50 μm to 300 μm. We chose STIN waveguides with 5 different diameters between 5 and 7 μm, which provide high enough $LP_{01}$ confinement while remaining single mode. After propagation we fitted the output intensity to a Gaussian and discarded the background. Figure 7(a) depicts the propagation losses for an injection NA of $\simeq 0.12$. We linearly fit the average losses of all diameters for every length value and the resulting dependency has a slope of -6.2 dB/mm. The small variation between propagation losses of different diameters indicates that scattering from side walls is not the dominant loss mechanism in our STIN waveguides. Noteworthy, Fig. 7(a) includes injection losses of ~-0.5 dB. Total losses for STIN and GRIN waveguides are very similar and within ~1 dB difference only. For the waveguides depicted in Figs. 6(b) and 6(c), the latter has an excess loss of 0.7 dB with respect to the STIN waveguide combining both injection as well as propagation losses. While low for a first proof of concept demonstration, losses are around one order of magnitude above the material's bulk absorption [22], but a factor of 3 lower than for 3D printed waveguides [6].

Figure 7(b) shows the evanescent coupling rate between neighboring waveguides. For our characterization we printed pairs of 300 µm-length and 7 µm-diameter waveguides with separations ranging from 0.5 µm up to 6 µm. Optical crosstalk between adjacent waveguides is based on the overlap of their respective evanescent fields, which decay exponentially over distance. Our evanescent coupling rate follows the expected exponential dependency, and the coupling rate reduces as $-1.32$ µm$^{-1}$ with waveguide separation. STIN waveguides in direct contact couple at a rate of 3 mm$^{-1}$, while this coupling drops to 0.02 mm$^{-1}$ for an intra-waveguide distance of 6 µm. The rapidly decaying evanescent coupling therefore allows a high integration density, supporting circuits with ~6000 waveguides per mm$^2$.

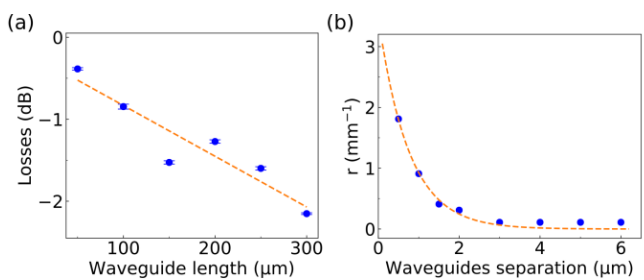

**Fig. 7.** (a) Propagation losses. (b) Evanescent-coupling rate between waveguide pairs as a function of waveguides separation. The optical injection NA is ≃0.12. The corresponding linear and exponential fits are depicted as dashed lines in both panels.

## 5. CONCLUSIONS

In summary, we have introduced a single-step process to manufacture photonic GRIN elements based on a commercial laser writing system. We first carried out different fundamental characterizations that independently confirmed the high level of control over a complex 3D refractive index distribution. Comparing these independent measurements (rectangular prisms with volume gratings) reveals, a very small $\Delta n$ deviation of 2% only ($\Delta n = 5 \cdot 10^{-3}$ compared to $\Delta n = 5.1 \cdot 10^{-3}$, respectively). This shows the excellent spatial and refractive index resolution of our technique. We then demonstrated (3+1)D printed GRIN volume holograms, as well photonic waveguides with a controlled number of propagating modes.

For the STIN and GRIN photonic waveguides, we determined the waveguides' NA by carefully fitting their propagation parameters under single-mode condition, characterized propagation losses and the evanescent coupling rates between neighboring waveguides. We find higher NA and mode confinement for GRIN waveguides. Moreover, significantly larger NAs should be achievable in the future, as the investigated and other commercial resins allow for larger refractive index modifications [19]. Further future efforts should reduce propagation losses in order to bring them closer to this of integrated silicon photonics. For the volume holograms, the results demonstrate printing an index distribution arbitrarily varying in 3D. Since currently the printed volumes are small, only a limited number of holograms is supported, and the fabrication of greater volumes to store more information is an important future objective. The currently employed printing technique would require stitching of different blocks to reach larger volumes since the field of view of the writing objective is limited. Any shift due to a stitching error would make different parts of the hologram out of phase. Thus, a thorough optimization of the fabrication process for stitching is necessary to approach mm sizes. One of the main goals of the future study is to fabricate a volume where many holograms are multiplexed and compare the diffraction efficiency with optically recorded counterparts. Moreover, over-modulated volume gratings fabricated by (3+1)D printing is another interesting avenue to explore, which requires a careful inspection [47]. For the characterization and calibration process, more recent and sensitive techniques can be adopted for high sensitivity demanding applications [48].

In general, the additive nature of our approach is a crucial asset. It makes the process less dependent on the working distance of microscope objectives, and just recently millimeter-sized photonic components utilizing the same TPP process have been demonstrated [7]. Furthermore, (3+1)D additive photonic fabrication has the potential to functionalize integrated photonic or electronic circuits, for example by adding scalable photonic interconnects [5,6] to bring large scale parallel communication to classical photonic or electronic chips. (3+1)D direct laser writing hence provides a highly versatile addition to the photonics toolbox.

**Funding.** European Union's Horizon 2020 (713694); Volkswagen Foundation (NeuroQNet II); Agence Nationale de la Recherche (ANR-17-EURE-0002, ANR-15-IDEX-0003); Region Bourgogne Franche-Comté; French Investissements d'Avenir; French RENATECH network and FEMTO-ST technological facility. Swiss National Science Foundation (514481).

**Disclosures.** The authors declare no conflicts of interest.

**Acknowledgments.** The two first authors, tagged with †, have contributed equally to the manuscript and the order of their names has been chosen randomly.